\documentstyle[12pt]{article}
\advance\textheight by 60pt
\advance\voffset by -40pt
\advance\textwidth by 50pt
\advance\oddsidemargin by -25pt
\advance\evensidemargin by -25pt

\def\single_space{\baselineskip 12pt plus 1pt minus 1pt}
\def\one_and_a_half_space{\baselineskip 19pt plus 1pt minus 1pt}
\def\double_spacesp{\baselineskip 25pt plus 2pt minus 2pt}

\def\atversim#1#2{\lower0.7ex\vbox{\baselineskip\zatskip\lineskip\zatskip
  \lineskiplimit 0pt\ialign{$\matth#1\hfil##\hfil$\crcr#2\crcr\sim\crcr}}}

\begin{document}
\begin{titlepage}
\begin{flushright}
{\bf
MAD/PH/745\\
SNUTP 93-6\\
YUMS 93-1\\
March 1993\\
}
\end{flushright}
\vskip 1.5cm
{\Large
{\bf
\begin{center}
Associated $J/\psi + \gamma$ production as a probe of the \\
polarized gluon distribution \\
\end{center}
}}
\vskip 1.0cm
\begin{center}
M. A. Doncheski\\
Department of Physics\\
University of Wisconsin USA\\
Madison, WI 53706\\
\vskip .3cm
and\\
\vskip .3cm
C. S. Kim \\
Department of Physics\\
Yonsei University\\
Seoul 120, Korea\\
\end{center}
\vskip 1.0cm
\begin{abstract}

Associated production of $J/\psi$ and a $\gamma$ has recently been
proposed as clean probe of the gluon distribution.  The same mechanism
can be used to probe the polarized gluon content of the proton in
polarized proton-proton collisions.  We study $J/\psi + \gamma$
production at both polarized fixed target and polarized collider energies.
\end{abstract}
\end{titlepage}
\double_spacesp

Interest in high energy spin physics has been recently revived with the
result from (and interpretations thereof) the EMC
collaboration\cite{ashman} on polarized $\mu-p$ scattering.  Processes in
polarized $pp$ collisions (such as achievable at an upgraded Fermilab
fixed target facility or at a polarized collider \cite{workshop})
sensitive to the polarized gluon content of the proton, such as
jets\cite{kunszt,guillet,sofferrhic}, direct
photons\cite{sofferrhic,qiu,was}, and heavy quark
production\cite{contogouris}, have been discussed.  Another intriguing
suggestion, due to Cortes and Pire\cite{cortes}, is to consider
$\chi_2(c\overline{c})$ production where the dominant lowest-order
subprocess would be $gg \rightarrow \chi_2$.  The partonic level
asymmetries for $\chi_{2}/\chi_{0}$ production have been calculated in
the context of potential models\cite{doncheski} and are large.  Low
transverse momentum quarkonium production in polarized $pp$ collisions
using other methods has also been considered\cite{contogouris,hidaka} as
has high $p_{_T}$ $\psi$ production \cite{robinett}.

In all cases of charmonium production, the experimental signal is
$\ell^+ \ell^-$ ($\ell = e$ or $\mu$) with the lepton-lepton invariant
mass giving the $J/\psi$ mass, since $\chi_J$ can decay radiatively to
$J/\psi + \gamma$, and the $J/\psi$ signature is quite clean.  As has
been noted\cite{madnrr}, the question of extracting the gluon
distribution is made less clean by the multitude of contributing
processes, {\it e.g.}:
\begin{eqnarray}
g + g & \to & \chi_{0,2} \nonumber \\
g + g & \to & \chi_J + g \nonumber \\
q + g & \to & \chi_{0,2} + q \nonumber \\
q + \bar{q} & \to & \chi_{0,2} + g \nonumber \\
g + g & \to & J/\psi + g \nonumber \\
g + g & \to & b(\to J/\psi + X) + \bar{b} \nonumber \\
q + \bar{q} & \to & b(\to J/\psi + X) + \bar{b}.
\end{eqnarray}
The simplicity of the Cortes and Pire idea is now gone.  A full
${\cal O}(\alpha_s^3)$ calculation of the spin-dependent production of
$\chi_J$ is necessary.  At low $p_{_T}$, $\chi_J$ production will also
involve $q + g$ and $q + \bar{q}$ initial states, while at high $p_{_T}$
in addition the $2 \to 2$ kinematics make the extraction of parton
distribution functions less direct.  Furthermore, a very careful
calculation is required because even processes with small cross section
can have a large effect on the asymmetry.  The extraction of
$\Delta g(x,Q^2)$ using inclusive $J/\psi$ will be a challenge.

Recently, $J/\psi$ produced in association with a $\gamma$ has been
proposed as a clean channel to study the gluon distribution at hadron
colliders\cite{kimdrees}.  The radiative $\chi_J$ decays can produce
$J/\psi$ at both low and high $p_{_T}$, but the photon produced will be
soft ($E \sim {\cal O}(400\;{\rm MeV})$).  If we insist that the
experimental signature consist of a $J/\psi$ and $\gamma$, with large but
equal and opposite $p_{_T}$ there is only one production mechanism
\begin{equation}
g + g \to J/\psi + \gamma.
\end{equation}
Following Ref.~\cite{kimdrees}, this mechanism has been proposed in
Ref.~\cite{sridhar2} to study the polarized gluon distribution in
polarized fixed target experiments; we perform a more detailed analysis,
including the analysis of this mechanism at the Brookhaven Relativistic
Heavy Ion Collider (RHIC) at both 50~GeV and 500~GeV center of mass
energy and at the Superconducting Super Collider (SSC).  Polarized
proton-proton operation is being considered for RHIC, for at least
several months data collection, while the tunnel design of the SSC has
been modified for the possible future inclusion of the Siberian Snakes
needed for polarized proton-proton mode.  Also, we list the full set of
helicity amplitudes for this process, explicitly stating the Lorentz
frame in which the $J/\psi$ helicities are given.

The full helicity amplitudes for $g + g \to J/\psi + \gamma$ can be
calculated following the approach of Gastmans, Troost and Wu\cite{GTW},
with the addition of explicit helicity polarization vectors for the
$J/\psi$.  A convenient set of polarization vectors can be found in
B\"ohm and Sack\cite{BS}.  These polarization vectors reduce to the
usual massive vector boson (+,$-$,0) polarization vectors in the parton
center of mass frame, and so, although the expressions for the helicity
amplitudes have Lorentz invariant form, the (+,$-$,0) only refer to
the $J/\psi$ helicity in this one particular frame.  We find only one
independent helicity amplitude ($M(++,++)$, where the `++,++' refer to
the helicity of $g_1 g_2,\gamma J/\psi$ respectively), and the remaining
5 non-zero helicity amplitudes can be found by crossing and parity
symmetries:
\begin{eqnarray}
M(++,++) & = & M(--,--) = C \frac{\hat{s} (\hat{s} - M^2)}
{(\hat{s} - M^2) (\hat{t} - M^2) (\hat{u} - M^2)} \nonumber \\
M(+-,-+) & = & M(-+,+-) = C \frac{\hat{u} (\hat{u} - M^2)}
{(\hat{s} - M^2) (\hat{t} - M^2) (\hat{u} - M^2)} \nonumber \\
M(-+,-+) & = & M(+-,+-) = C \frac{\hat{t} (\hat{t} - M^2)}
{(\hat{s} - M^2) (\hat{t} - M^2) (\hat{u} - M^2)}
\end{eqnarray}
where $C = \frac{\mbox{$4 e_q e g_s^2 R(0) M \delta^{ab}$}}
{\mbox{$\sqrt{3 \pi M}$}}$.  Here, $M$ is the $J/\psi$ mass, $\hat{s}$,
$\hat{t}$ and $\hat{u}$ are the usual Mandelstam variables, $R(0)$ is the
radial wavefunction at the origin of the $c \bar{c}$ in the $J/\psi$ and
$a,b$ are the color indices of the incident gluons.  Thus, the (spin and
color) summed and averaged matrix element squared can be
found\cite{kimdrees}:
\begin{eqnarray}
\overline{|M(g + g \to J/\psi + \gamma)|^2} & = &
\frac{(16 \pi)^2 \alpha \alpha_s^2 M |R(0)|^2}{27}
\left[\frac{\hat{s}^2}{(\hat{t} - M^2)^2 (\hat{u} - M^2)^2} \right.
\nonumber \\
& + & \left. \frac{\hat{t}^2}{(\hat{u} - M^2)^2 (\hat{s} - M^2)^2}
+ \frac{\hat{u}^2}{(\hat{s} - M^2)^2 (\hat{t} - M^2)^2} \right].
\end{eqnarray}
$|R(0)|^2$ can be related to the leptonic width of the $J/\psi$:
\begin{eqnarray}
\Gamma(J/\psi \to e^+ e^-) & = & \frac{16 \alpha^2}{9 M^2} |R(0)|^2
= 4.72 \; {\rm keV} \nonumber \\
|R(0)|^2 & = & 0.48 \; {\rm GeV}^3.
\end{eqnarray}

We are interested in the longitudinal spin-spin asymmetry, defined as:
\begin{equation}
A_{LL} = \frac{\sigma(++) - \sigma(+-)}{\sigma(++) + \sigma(+-)}
\end{equation}
where $\sigma(++)$ ($\sigma(+-)$) is the cross section for the collision
of 2 protons with the same (opposite) helicities.  This can be calculated
in the parton model,
\begin{equation}
A_{LL} \sigma = \int dx_1 \; dx_2 \; \hat{a}_{LL} \; \hat{\sigma} \;
\Delta g(x_1,Q^2) \; \Delta g(x_2,Q^2)
\end{equation}
where $\hat{\sigma}$ is the parton level cross section (related to
$\overline{|M|^2}$ given earlier), $\Delta g(x,Q^2)$ is the polarized
gluon distribution in the proton ($ = (g^+(x,Q^2) - g^-(x,Q^2))$ where
$g^+(x,Q^2)$ ($g^-(x,Q^2)$) is the distribution for gluons with the
same (opposite) helicity as that of the proton) and $\hat{a}_{LL}$ is
the parton level asymmetry
\begin{equation}
\hat{a}_{LL} = \frac{\hat{\sigma}(++) - \hat{\sigma}(+-)}
                    {\hat{\sigma}(++) + \hat{\sigma}(+-)}.
\end{equation}
Given the known helicity amplitudes for this process, the parton level
asymmetry is simply
\begin{equation}
\hat{a}_{LL} = \frac{\hat{s}^2 (\hat{s} - M^2)^2 - \hat{t}^2
(\hat{t} - M^2)^2 - \hat{u}^2 (\hat{u} - M^2)^2}
{\hat{s}^2 (\hat{s} - M^2)^2 + \hat{t}^2 (\hat{t} - M^2)^2
+ \hat{u}^2 (\hat{u} - M^2)^2}.
\end{equation}

Measurable quantities of interest are the $p_{_T}$ distribution and the
joint $p_{_T}$---$y_1$---$y_2$ distribution with $y_1 = y_2 =0$, where
$y_{1(2)}$ is the rapidity of the $\gamma$ ($J/\psi$).  In the latter
case, both partons have the same Bjorken-$x$ (which is a function of
$p_{_T}$ only).  The corresponding asymmetries are given by:
\begin{eqnarray}
A^1_{LL} &=& \frac{\sigma(++) - \sigma(+-)}
             {\sigma(++) + \sigma(+-)} \nonumber \\
A^2_{LL} &=& \frac{\frac{\mbox{$d \sigma(++)$}}{\mbox{$dp_{_T}$}}
          - \frac{\mbox{$d \sigma(+-)$}}{\mbox{$dp_{_T}$}}}
           {\frac{\mbox{$d \sigma(++)$}}{\mbox{$dp_{_T}$}}
          + \frac{\mbox{$d \sigma(+-)$}}{\mbox{$dp_{_T}$}}}
           \nonumber \\
A^3_{LL} &=& \frac{\frac{\mbox{$d \sigma(++)$}}
                  {\mbox{$dp_{_T} dy_1 dy_2$}}|_{y_1=y_2=0}
           - \frac{\mbox{$d \sigma(+-)$}}
                  {\mbox{$dp_{_T} dy_1 dy_2$}}|_{y_1=y_2=0} }
                  {\frac{\mbox{$d \sigma(++)$}}
                  {\mbox{$dp_{_T} dy_1 dy_2$}}|_{y_1=y_2=0}
           + \frac{\mbox{$d \sigma(+-)$}}
                  {\mbox{$dp_{_T} dy_1 dy_2$}}|_{y_1=y_2=0} }.
\end{eqnarray}
Note that $A^3_{LL}$ is proportional to $[\Delta g(x(p_{_T}),Q^2)]^2$.
Another interesting theoretical concept (though not measurable
expertimentally) is the average $\hat{a}_{LL}$, or `resolving power'.  It
is defined in the following way
\begin{equation}
\langle \hat{a}_{LL} \rangle \sigma = \int dx_1 \; dx_2 \; \hat{a}_{LL}
\; \hat{\sigma} \; g(x_1,Q^2) \; g(x_2,Q^2).
\end{equation}
As we wish to determine if a given experimental scenario can shed light
on the size of the polarized gluon in the proton, we need, in addition to
calculating the asymmetry, to estimate the experimental uncertainty in
the asymmetry.  We will approximate the uncertainty by the statistical
uncertainty, since ratios of cross sections should be relatively free of
systematic uncertainties.  The statistical uncertainty in the measurement
of an asymmetry is given by $\delta A$, where
\begin{equation}
\delta A = \frac{\sqrt{1 - A^2}}{\sqrt{N}}
\end{equation}
and $N$ is the number of events.

We examine this process in several different experimental settings.
First, we consider an hypothetical fixed target experiment and to be
specific, take the proton beam energy to be 800~GeV (such as would exist
at the upgraded Fermilab fixed target facility).  In order to estimate
the luminosity possible at such an experiment, we must make some
assumptions.  First, the Main Injector at Fermilab can provide
$\sim 10^{14}$~(unpolarized protons)/sec, with a 65\% duty
cycle\cite{MI}.  We'll assume a one month run, at a much reduced proton
rate (say, a factor of 100), combined with a small polarized gas ($H_2$)
jet target (approximately 1~cm long).  This will give, we think, a very
conservative estimate of $\int {\cal L} dt = 50\; {\rm pb}^{-1}$.  We
place no cuts on the rapidity of the photon or $J/\psi$, nor on the
$p_{_T}$ of the photon or leptons.  We find a cross section of
approximately 200~pb, most of which is at low $p_{_T}$.  The resolving
power (or average $\hat{a}_{LL}$) is found to be about 28\%.  We use the
polarized distributions of Bourrely, Guillet and Chiappetta\cite{BGC}.
They provide 2 sets of distributions, one with a large polarized gluon
distribution and small polarized strange quark distribution (we'll refer
to it as the set BGC0) and one with a moderately large polarized gluon
and moderately large polarized strange quark distribution (we'll refer to
this set as BGC1).  The $p_{_T}$ distribution is shown in Figure 1a (in
cross section) and in Figure 1b (in $A^2_{LL}$).  We were also interested
the asymmetry $A^3_{LL}$, (technically, instead of taking $y_1$ and $y_2$
derivatives, we bin the events in the usual way, displaying the contents
of the bin with $-0.1 \leq y_1,y_2 \leq 0.1$).  The results are shown in
Figure~3a (distribution in cross section) and 3b ($A^3_{LL}$ {\it vs.}
$p_{_T}$).  We present in Table~1 the total number of events expected (at
all $p_{_T}$ and $y_{1,2}$ consistent with our cuts) as well as the
`resolving power' and asymmetry $A^1_{LL}$ and an estimate of the
statistical uncertainty, $\delta A^1_{LL}$.  We also list the number of
events in a single $p_{_T}$ bin ($p_{_T}$ given in the table caption),
and $A^2_{LL}$ and $\delta A^2_{LL}$ for that particular $p_{_T}$ bin.
Finally, we present the the number of events in the same $p_{_T}$ bin,
further restricting the events to lie within $|y_{1,2}| \leq 0.1$, and
the value of $A^3_{LL}$ and $\delta A^3_{LL}$ in the particular $p_{_T}$
bin.  These are representative results.  Higher statistics can be
obtained by the inclusion of all $p_{_T}$ bins.

At this point, we would like to further address the work of
Ref.~\cite{sridhar2}.  The large asymmetries shown are surprising, and in
our opinion not correct.  The parton level asymmetry, making the
following replacements for $\hat{t}$ and $\hat{u}$ ({\it i.e.} working in
the parton center of mass frame):
\begin{eqnarray}
\hat{t} & = & -\frac{1}{2} (\hat{s} - M^2) (1 - \cos \theta) \nonumber \\
\hat{u} & = & -\frac{1}{2} (\hat{s} - M^2) (1 + \cos \theta)
\end{eqnarray}
reduces to
\begin{equation}
\hat{a}_{LL} = \frac{1 -\frac{1}{8}[(1 + 6 \cos^2 \theta + \cos^4 \theta)
+ \frac{2 M^2}{\hat{s}}(1 - \cos^4 \theta)
+ \frac{M^4}{\hat{s}^2}(1 - \cos^2 \theta)^2]}
{1 + \frac{1}{8}[(1 + 6 \cos^2 \theta + \cos^4 \theta)
+ \frac{2 M^2}{\hat{s}}(1 - \cos^4 \theta)
+ \frac{M^4}{\hat{s}^2}(1 - \cos^2 \theta)^2]}.
\end{equation}
Here $\cos \theta$ is measured in the parton center of mass frame.  It is
obvious that for $\cos \theta = \pm 1$, $\hat{a}_{LL}$ is a minimum
(actually zero), and so, for any $\hat{s}$, the maximum of $\hat{a}_{LL}$
should be at $\cos \theta = 0$.  In this limit, the asymmetry reduces to
\begin{equation}
\hat{a}_{LL}(\cos \theta = 0) = \frac{1 - \frac{1}{8} \left(
\frac{\mbox{$\hat{s} + M^2$}}{\mbox{$\hat{s}$}} \right)^2}
{1 + \frac{1}{8} \left( \frac{\mbox{$\hat{s} + M^2$}}{\mbox{$\hat{s}$}}
\right)^2}.
\end{equation}
Two further limiting cases are possible, namely production at threshold
($\hat{s} = M^2$) which gives $\hat{a}_{LL} = \frac{1}{3}$ and production
at very high energy ($\hat{s} \to \infty$) which gives $\hat{a}_{LL} =
\frac{7}{9}$.  For $\sqrt{\hat{s}} = \sqrt{s} = 38.75$~GeV (the fixed
target energy considered both here and in Ref.~\cite{sridhar2}), the
parton level asymmetry is near it's maximum value.  Since
$\Delta g(x,Q^2)/g(x,Q^2) \leq 1$ generally, the maximum observable
asymmetry is bounded by the maximum parton level asymmetry.  Thus we are
unable to understand the prediction, in Ref.~\cite{sridhar2}, that the
observable asymmetry can be as large as 85\%.

Next, we consider collider experiments at RHIC.  RHIC is a high
luminosity (${\cal L} = 2 \times 10^{32} \; {\rm cm}^{-2} {\rm sec}^{-1}
= 6000 \; {\rm pb}^{-1}/{\rm yr}$) collider capable of producing proton
on proton collisions for center of mass energies between 50 and 500~GeV.
A program of polarized proton on proton collisions, at full energy and
luminosity, is being discussed\cite{RSC}.  We will assume a nominal
running time of 2 months, at full luminosity, for 50~GeV and 500~GeV
each.  In order to be somewhat conservative, we will estimate event
numbers based on 300~pb$^{-1}$ integrated luminosity.  We will assume a
generic collider type detector, and in order to simulate the acceptance
we will require the photon and electrons observed to lie in the rapidity
range $|y| \leq 2$ (this simulates the acceptance of the proposed STAR
detector at RHIC\cite{STAR}, level 2 for photons and electrons.  We will
not consider the possibility of the detection of the $\mu^+ \mu^-$ final
state at RHIC).  Furthermore, we will (rather arbitrarily) require the
$p_{_T}$ of the photon larger than 1~GeV in the following discussion.  We
present our results for the $p_{_T}$ distribution in Figure 3a, and
$A^2_{LL}$ in Figures 3b ($\sqrt{s} = 50$~GeV) and 3c
($\sqrt{s} = 500$~GeV).  See Figure 4a for
$\frac{\mbox{$d \sigma$}}{\mbox{$dp_{_T} dy_1 dy_2$}}$ {\it vs.} $p_{_T}$
and Figures 4b ($\sqrt{s} = 50$~GeV) and 4c ($\sqrt{s} = 500$~GeV) for
$A^3_{LL}$ {\it vs.} $p_{_T}$.  The `resolving power' increases with
energy (actually $p_{_T}$), even though the observed asymmetry
decreases.  This is simply a consequence of the behavior of the polarized
gluon distribution.  Please refer to Table~1 for some representative
results.

Finally, we consider a collider experiment at the SSC.  The luminosity of
the SSC is ${\cal L} = 10^{33} \; {\rm cm}^{-2} {\rm sec}^{-1} =
30000 \; {\rm pb}^{-1}/{\rm yr}$.  We will again assume a running time of
2 months at full luminosity and energy, and conservatively calculate
event numbers based on 1500~pb$^{-1}$ integrated luminosity.  We require
the photons and leptons to have $p_{_T} \geq 10$~GeV and lie in the range
$|y| \leq 2.5$ (these approximate the acceptances of the SDC
detector\cite{TDR}).  In this case, the resolving power is quite high,
$\langle \hat{a}_{LL} \rangle = 60\%$, although because of the extremely
small-$x$ probed the observed asymmetry $A^1_{LL}$ is tiny.  Similarly,
$A^2_{LL}$ and $A^3_{LL}$ are both smaller than 1\% for all
$p_{_T} < 125$~GeV, while there will only be a handful of events at (or
beyond) $p_{_T} \sim 25$~GeV, so there is no observable asymmetry.
Again, see Table~1 for some representative results.

In conclusion, we have studied the process
$p + p \to J/\psi + \gamma + X$ in polarized proton-proton collisions.
We first presented the necessary helicity amplitudes and discussed the
calculation.  Then we studied this process at polarized fixed target and
in colliders, at polarized RHIC (50 and 500~GeV center of mass energy)
and at polarized SSC.  Our results indicate that a polarized
(double spin) fixed target program can be very useful in the
determination of the polarized gluon distribution.  It is unfortunate
that no such experiment is planned.  RHIC (especially at lower energies)
is an excellent probe of the polarized gluon distribution.  Since
$A^3_{LL}$ is directly proportional to
$[\Delta g(x(p_{_T}),Q^2)/g(x(p_{_T}),Q^2)]^2$, this distribution
provides an easy determination of the polarized gluon distribution at
various $x$ values.  It will prove especially useful to measure this
distribution at several center of mass energies.  Even a measurement of
$A^2_{LL}$ can provide much useful information (though it is not clear
whether the higher statistics involved in this measurement will outweight
the cleanliness of the extraction of the polarized gluon distribution in
a measurement of $A^3_{LL}$).  The SSC probes a much lower $x$ in this
process, and since $\Delta g(x,Q^2)/g(x,Q^2) \ll 1$ there is no
measurable asymmetry.  However, the `resolving power' at SSC is still
very large, so the smallness of the asymmetry is purely a consequence of
the small-$x$ behavior of $\Delta g(x,Q^2)$.  Polarized SSC can still be
a useful tool for the study of high energy spin properties of the proton
by utilizing a subprocess that will probe larger $x$ ({\it e.g.} heavy
Higgs production).  We should also point out that we have considered only
the color singlet model of heavy quarkonium production in this paper.  A
similar analysis can be performed using local duality, if it is
determined at HERA that this mechanism contributes to $J/\psi + \gamma$
production\cite{kimreya}.  Some slight modifications will be required,
namely the inclusion of charm in the proton (this effect should be small)
and light $q\bar{q}$ fusion, and in addition the modification of the
parton level asymmetries.  As a final related comment, we plan to study
$J/\psi + \gamma$ production at HERA using a polarized lepton beam and
angular distributions of the final leptons to learn something of the
polarized gluon distribution of the photon.

\vspace*{0.4cm}
{\Large{\bf \noindent  Acknowledgements}}
\vspace*{0.4cm}

The work of MAD was supported in part by the U.~S. Department of Energy
under Contract No.~DE-AC02-76ER00881, in part by the Texas National
Research Laboratory Commission under Grant Nos.~RGFY9173 and RGFY9273,
and in part by the University of Wisconsin Research Committee with funds
granted by the Wisconsin Alumni Research Foundation.  The work of CSK was
supported in part by the Korean Science and Engineering Foundation, in
part by the Center of Theoretical Physics at Seoul National University
and in part by a Yosei University Faculty Research Grant.  The authors
would like to thank V. Barger, H. Chehime, T. Rizzo, A. Yokosawa and D.
Zeppenfeld for useful discussions, and especially to R. Robinett for
critically reading this manuscript.

\newpage

\begin{table}[h]
\begin{tabular}{c|c|c|c|c|c|c|c}  \hline \hline
         & $N_{TOT}$ & $\langle \hat{a}_{LL} \rangle$
          & $A^1_{LL}(\delta A^1_{LL})$ & $N_{p_{_T}}$
          & $A^2_{LL}(\delta A^2_{LL})$ & $N_{p_{_T}}$
          & $A^3_{LL}(\delta A^3_{LL})$   \\
         &           &
          &                             &
          &                             &$|y_{1,2}| \leq 0.1$
          & \\ \hline \hline
Fixed        & 10500 & 28.4\% & 12.5\% (1\%)  & 5000 & 16\% (1.4\%)
&  200                & 22\% (6\%)   \\
Target       &       &        &  3.2\% (1\%)  &      &  4\% (1.4\%)
&                     &  5\% (6\%)   \\ \hline
RHIC         & 11430 & 43.3\% & 19.1\% (1\%)  & 4500 & 26\% (1.5\%)
& 1080                & 32\% (3\%)   \\
50 GeV       &       &        &  4.6\% (1\%)  &      &  8\% (1.5\%)
&                     &  8\% (3\%)   \\ \hline
RHIC         & 86400 & 44.7\% &  .4\% (0.3\%) & 4500 & 1.7\% (1.5\%)
&  840                & 1.8\% (3\%)  \\
500 GeV      &       &        & .05\% (0.3\%) &      & .2\% (1.5\%)
&                     & .3\% (3\%)   \\ \hline
SSC          &  8835 & 60.2\% &  .005\% (1\%) & 3000 & .008\% (2\%)
&  540                & .01\% (4\%)  \\
             &       &        & .0006\% (1\%) &      & .001\% (2\%)
&                     & .001\% (4\%) \\ \hline
\end{tabular}
\caption{Summary of representative predictions for $J/\psi + \gamma$
production in polarized proton-proton interactions.  $N_{TOT}$ is the
total number of events above some minimum $p_{_T}$ (= 0~GeV for fixed
target, 1~GeV for RHIC and 10~GeV for SSC).
$\langle \hat{a}_{LL} \rangle$ is the `resolving power' as defined in the
text (this is independent of the polarized parton distributions).
$A^i_{LL}$ and $\delta A^i_{LL}$ are defined in the text; the upper entry
corresponds to the large $\Delta g(x,Q^2)$ (set BGC0) and the lower entry
corresponds to the moderately large $\Delta g(x,Q^2)$ (set BGC1).
$N_{p_{_T}}$ is the number of events in the particular $p_{_T}$ bin
(0.5-1.5~GeV for fixed target, 1-2~GeV for RHIC at 50~GeV, 3-5~GeV for
RHIC at 500~GeV and 10-20~GeV for SSC).}

\end{table}

\newpage

\vspace*{0.4cm}
{\Large{\bf \noindent  Figure Captions}} \\
\vspace*{0.4cm}

Figure 1 - $p_{_T}$ distribution, $\frac{\mbox{$d \sigma$}}
{\mbox{$dp_{_T}$}}$ {\it vs.} $p_{_T}$ (1a) and $A^2_{LL}$ {\it vs.}
$p_{_T}$ (1b) for large $\Delta g(x,Q^2)$ (solid line) and for moderately
large $\Delta g(x,Q^2)$ (dashed line) at fixed target. \\

Figure 2 - $\frac{\mbox{$d \sigma$}}
{\mbox{$dp_{_T} dy_1 dy_2$}}|_{y_1 = y_2 = 0}$ {\it vs.} $p_{_T}$ (2a)
and $A^3_{LL}$ {\it vs.} $p_{_T}$ (2b) for large $\Delta g(x,Q^2)$
(solid line) and moderately large $\Delta g(x,Q^2)$ (dashed line) at
fixed target. \\

Figure 3 - $p_{_T}$ distribution, $\frac{\mbox{$d \sigma$}}
{\mbox{$dp_{_T}$}}$ {\it vs.} $p_{_T}$ (3a) for RHIC at
$\sqrt{s} = 500$~GeV (solid line) and at $\sqrt{s} = 50$~GeV (dot-dashed
line), and $A^2_{LL}$ {\it vs.} $p_{_T}$ for RHIC at $\sqrt{s} = 50$~GeV
(3b) and at $\sqrt{s} = 500$~GeV (3c) for large $\Delta g(x,Q^2)$ (solid
line) and for moderately large $\Delta g(x,Q^2)$ (dashed line). \\

Figure 4 - $\frac{\mbox{$d \sigma$}}
{\mbox{$dp_{_T} dy_1 dy_2$}}|_{y_1 = y_2 = 0}$ {\it vs.} $p_{_T}$ (4a)
for RHIC at $\sqrt{s} = 500$~GeV (solid line) and at $\sqrt{s} = 50$~GeV
(dot-dashed line), and $A^3_{LL}$ {\it vs.} $p_{_T}$ for RHIC at
$\sqrt{s} = 50$~GeV (4b) and at $\sqrt{s} = 500$~GeV (4c) for large
$\Delta g(x,Q^2)$ (solid line) and moderately large $\Delta g(x,Q^2)$
(dashed line). \\

\end{document}